\documentclass[letterpaper,12pt]{article}
\usepackage{amsmath}
\usepackage{subfig}
\usepackage[dvips]{graphicx}
\usepackage[dvips]{color}
\usepackage{cite}
\usepackage[thinspace,squaren]{SIunits}
\newcommand{\nm}{\nano\meter}

\newcommand{\RandJ}{\vec{\mathbf{J}}}

\newcommand{\RandL}{\mathbf{L}}

\newcommand{\RandP}{\vec{\mathbf{P}}}

\newcommand{\RandX}{\mathbf{X}}
\newcommand{\RandY}{\mathbf{Y}}

\newcommand{\RandTheta}{\mathbf{\Theta}}
\newcommand{\RandDX}{\mathbf{\Delta X}}
\newcommand{\RandDY}{\mathbf{\Delta Y}}
\begin{document}
\title{
A New Characterization of Fine Scale Diffusion on the Cell Membrane
\thanks{This work was supported in part by NIH grant
P50 GM085273, supporting the Center for Spatiotemporal
Modeling of Cell Signaling, and by NIH grants
R01 GM49814 and R01 AI051575.} 
}

\author{
    Flor A. Espinoza \\
    Department of Mathematics and Statistics \\
    Kennesaw State University \\
    Atlanta, GA 30144-5591 USA
\and
    Stanly L. Steinberg$^\dagger$\\
    The Center for the Spatiotemporal Modeling of Cell Signaling \\
    Department of Mathematics and Statistics \\
    University of New Mexico \\
    Albuquerque NM 87131-1141 USA
    }

\maketitle
\newpage \clearpage
\tableofcontents \listoftables \listoffigures

\newpage
\begin{abstract}

We use a large single particle tracking data set to analyze the short time
and small spatial scale motion of quantum dots labeling proteins in cell
membranes.  Our analysis focuses on the jumps which are the changes in the
position of the quantum dots between frames in a movie of their motion.
Previously we have shown that the directions of the jumps are uniformly
distributed and the jump lengths can be characterized by a double power
law distribution.

Here we show that the jumps over a small number of time steps can be
described by scalings of a {\em single} double power law distribution.
This provides additional strong evidence that the double power law
provides an accurate description of the fine scale motion.  This more
extensive analysis provides strong evidence that the double power law
is a novel stable distribution for the motion. This analysis provides
strong evidence that an earlier result that the motion can be modeled
as diffusion in a space of fractional dimension roughly 3/2 is correct.
The form of the power law distribution quantifies the excess of short
jumps in the data and provides an accurate characterization of the fine
scale diffusion and, in fact, this distribution gives an accurate
description of the jump lengths up to a few hundred nanometers.
Our results complement of the usual mean squared displacement analysis
used to study diffusion at larger scales where the proteins are more
likely to strongly interact with larger membrane structures.

\end{abstract}


\newpage \clearpage
\setcounter{equation}{0} \setcounter{figure}{0} \setcounter{table}{0}
\section{Introduction}

Our goal is to better understand the fine scale motion of proteins in cell
membranes. We do this by studying a large single particle tracking data
set for the $IgE$ high affinity receptor $Fc\epsilon RI$ tagged with quantum
dots (QDs) \cite{espinoza12,yinghzs09} which were taken from movies of the
positions of the QDs. Standard time series analysis requires the analysis
of the jumps which are the changes in the positions of the QDs between
frames in the movie as was done in \cite{espinoza12,yinghzs09}.  
The components of the jumps in a Brownian random walk in two dimensions
are normally distributed if and only if the directions of the jumps
are uniformly distributed and the lengths of the jumps have a chi, or
equivalently, Weibull distribution.  A main result of the previous
studies is that the directions of the jumps are uniformly distributed,
just as in Brownian motion, but the jump lengths do not have a
chi distribution.  Instead the studies provide evidence that the jumps
can be described using a double power law distribution.  It was also shown
in \cite{espinoza12}, that a double power law fits the jump data better
than general chi or general Weibull distributions.  Our power law
distribution has two power law behaviors, most importantly one for
small jumps and another less important power law behavior for large jumps.

Here we study the jumps over a few time steps, not just one time step,
and provide much stronger
evidence that the previous descriptions of the motion are accurate.
It came as a surprise to us that the double power laws for multiple time
steps are related to the one time step power law by a square root of time
scaling law, just as in Brownian motion.  This allows us to characterize
the diffusion on short time and small spatial scales using a single double
power law probability distribution function with scaling and is the
key to providing stronger evidence for the previous characterization
of the motion. Actually the double power law is surprisingly accurate
for moderate spatial and temporal scales, does not have a heavy tail
\cite{clausetsn09,nolan13} and, in fact, has many finite moments.

An important point is that the measured motion of the quantum dots has several
components one of which is the actual motion of the protein.  These include
movement of the dot while the image is being taken, noise in the imaging
equipment, and the super resolution imaging process.  These errors have been
studied extensively in the context of MSD analysis, see e.g.
\cite{slavendoyle2014,shuangBKWZ13,keptenbg2013,perssonlue2013,arnspangscwl2013,weber2012,monniergmhlb2012,nandiHL12,dascc2009}.
However, it is reasonable to assume that all these random processes
are independent, and all but the underlying motion of the protein
are normally distributed. If the underlying motion of the protein
was normally distributed, then the measured motion would also
be normally distributed, but it is not. The most likely and cause
of the double power law behavior is the motion of the proteins.

This work has important implications for modeling protein motion.
The Central Limit Theorem implies that if we could model our data
as an random walk in a homogeneous medium that has independent and
identically distributed (IID) jumps with double power law distributed
jump lengths, then the components of the jumps over many time steps
should approach a normal distribution, or equivalently, the jump angles
must be uniformly distributed and the jump lengths must have a chi
distribution.  We can view our data as sampling the motion after the
protein has made many smaller time steps. If the motion could be
described as an IID random walk then jump lengths must have a chi
distribution, which they do not have.  We view this to mean that the
diffusion of proteins in the cell membrane is a complex process that
cannot be accurately modeled using an IID random walk in a homogeneous
medium. We note that previous work \cite{espinoza12} supports the
assumption that the jumps are independent.
Collaborators have recently made simulations of random walks in
non-homogeneous mediums that have statistics similar to our data \cite{TCS}.

To understand why we view the motion as diffusion in a space
of dimension 3/2, one needs to know that in dimension $n$, IID random
walks where the components of the motion are normally distributed with
mean zero and standard deviation $\sigma$, the motion can also be
characterized as uniform on the unit sphere and radially by the chi
distribution with $n$ degrees of freedom and scale factor $s$,
$c(r,s,n) = c(r/s,n)/s$ where \cite{espinoza12}
\begin{equation}\label{Chi}
c(r,n) = \frac{2}{2^{n/2} \Gamma ({n/2})} \,
        r^{n-1 } \, e^{-\frac{r^2}{2}} \;,
\end{equation}
$r \geq 0$, $n \, s^2 = \sigma^2 >0$ and $n \geq 1$.
In two dimensions this is known as the chi or Weibull distribution
while in dimension three it is the Maxwell-Boltzmann distribution.
For small $r$, this distribution has the from
\[
c(r,n) \approx C \, r^{n-1}  \,,
\]
where $C$ is a constant. Consequently, for the fine scale motion, if a
probability distribution function for the lengths of jumps has the form
\[
p(r) \approx C \, r^{d-1}  \,,
\]
for small $r$ and where $C$ is a constant, then we say that the motion
can be modeled as being in a space of dimension $d$. For our data,
$d \approx 3/2$.  This is a quantitative measure of the restrictions
on the motion of the protein on small spatial and temporal scales.

Our results complement those obtained by mean squared displacement
analysis which involves using data from more time steps than used here
and consequently are appropriate to temporal and spatial scales where
the proteins may have strong interactions with other membrane structures
\cite{babanishida12,lidkelcl11,changrosenthal11,firstpassage11,wellslgpclww10,manley10,alcorga09,saxton08}.

The biology and microscopy involved in creating the data are described
in detail in \cite{andrewspmhdowl09,andrewslpbwo08},
so we will not repeat this here.  We will use the same notation
as in \cite{espinoza12}, which we now summarize.

\subsection{Random Variables}

We model the QDs positions using vector valued random
variables:
\[
\RandP_n = (\RandX_n, \RandY_n) \,,\quad 1 \leq n \leq N \,,\quad N > 0 \,,
\]
where $\RandX_n$ and $\RandY_n$ are real valued random variables
and $N$ and $n$ are integers.  The jumps are also random variables:
\[
\RandJ_n = \RandP_n - \RandP_{n-1} = (\RandDX_n, \RandDY_n)
\,,\quad 2 \leq n \leq N \,.
\]
In polar coordinates, the lengths of the jumps $\RandL_n$ and the angles
$\RandTheta_n$ between the jump vectors and the $x$-axis are also real
valued random variables:
\[
\RandL_n = \| \RandJ_n \| = \sqrt{\RandDX_n^2 + \RandDY_n^2}
	\,,\quad \RandTheta_n = \arctan(\RandDY_n,\RandDX_n) \,,\quad
	2 \leq n \leq N \,,
\]
where $\arctan$ gives a value in $(-\pi , \pi]$ such that if $\RandL_n \neq 0$,
then $\cos(\RandTheta_n) = \RandDX_n/\RandL_n$ and $\sin(\RandTheta_n) = \RandDY_n/\RandL_n$,
and consequently, $\tan(\RandTheta_n) = \RandDY_n/\RandDX_n$ if
$\RandDX_n \neq 0$. If $\RandJ = (0,0)$, then $\RandTheta = 0$ (in matlab).

\subsection{Tracking Data}

\begin{table}
\begin{center}
 \begin{tabular}{|r|rrr|r|}
\hline
    & $k = 1$ & $k = 2$ & $k = 3$ \\
\hline
$A$ & 407,669 & 346,750 & 302,256 \\
$B$ & 353,368 & 300,517 & 261,474\\
\hline
\end{tabular}
\caption{The total number of jumps for 1, 2, and 3 time steps that
are in data sets $A$ and $B$.  \label{TotalJumps}}
\end{center}
\end{table}

The analyzed data were taken from unstimulated cells where a subset
of the Fc$\epsilon$RI on the cell membrane were labeled with QDs.
We studied two large data sets called $A$ and $B$. We analyzed
each data set individually as it is useful to see the differences
between the two sets but for our new results we combine the data sets
to obtain better accuracy.
The data contain $M>0$ tracks with $N>0$ positions described by
\[
x_{m,n} \,,\quad y_{m,n} \,,\quad v_{m,n} \,,\quad
        1 \leq m \leq M \,,\quad 1 \leq n \leq N\,.
\]
The vectors
\[
\overrightarrow{r}_{m,n} = (x_{m,n}, y_{m,n})
\]
estimate the position of the QDs.  If $v_{m,n} = 1$, then the
QD is on and the position of the QD is valid data, while if
$v_{m,n} = 0$, the QD is off.

The length of a valid jump of over $k$ times steps is
\[
L_{m,n,k} = \| \overrightarrow{r}_{m,n+k}-\overrightarrow{r}_{m,n} \| \,,\quad
        1 \leq m \leq M \,,\quad 1 \leq n \leq N-k\,.
\]
The jump is valid provided that the QD is on
at all times $n$ through $n+k$, that is
\[
v_{m,n}*v_{m,n+1} * \cdots * v_{m,n+k} = 1 \,,\quad
        1 \leq m \leq M \,,\quad 1 \leq n \leq N-k\,.
\]
The number of valid jumps in data sets A and B is given in
Table \ref{TotalJumps}. These are large data sets.

\newpage \clearpage
\setcounter{equation}{0} \setcounter{figure}{0} \setcounter{table}{0}
\section{Analyzing the Data}

\begin{figure}[ht]
\centering
\begin{tabular}{cc}
\subfloat[Data set A]
{
\includegraphics[width=0.4\textwidth]{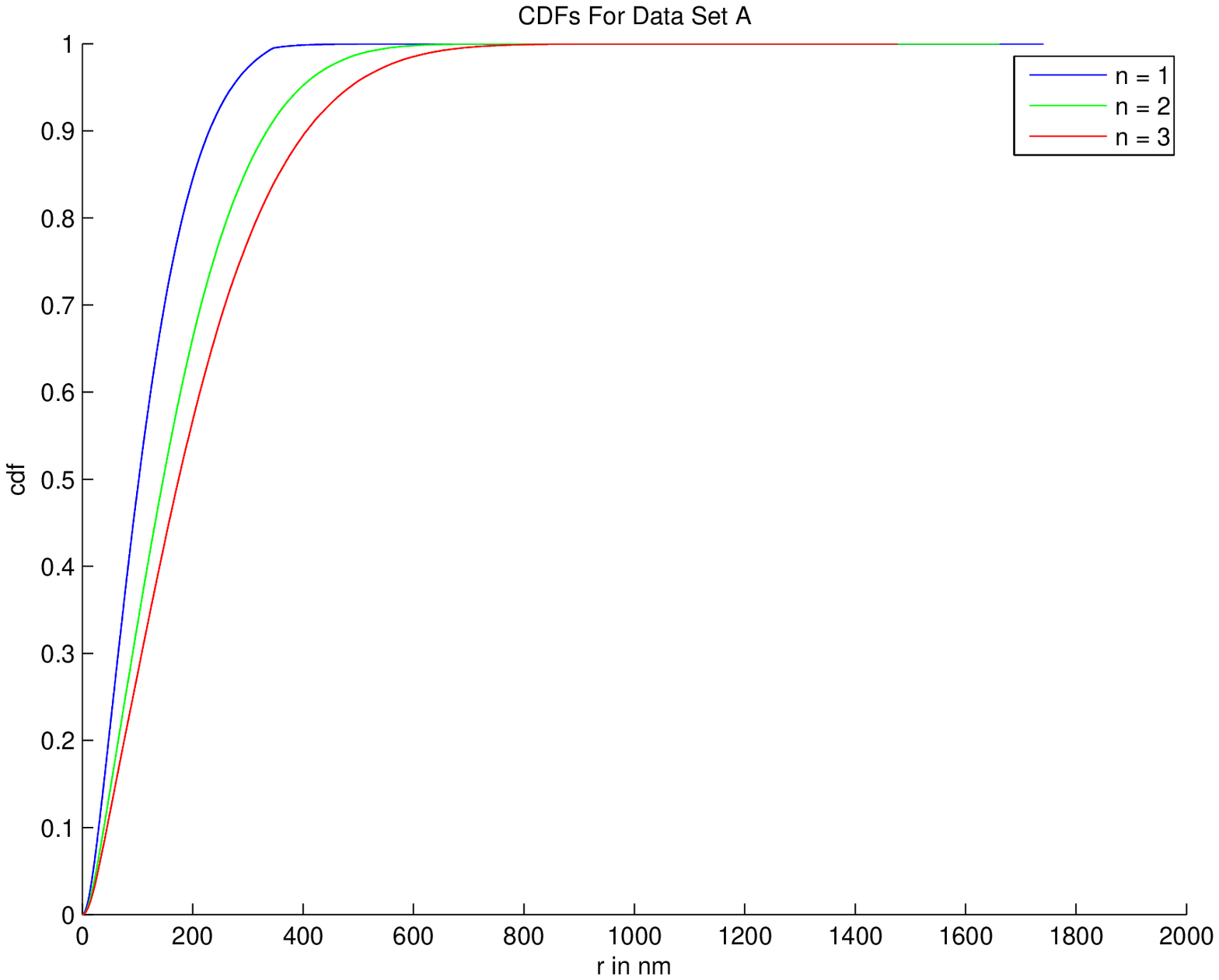}
} \qquad

&
\subfloat[Data set B]
{
\includegraphics[width=0.4\textwidth]{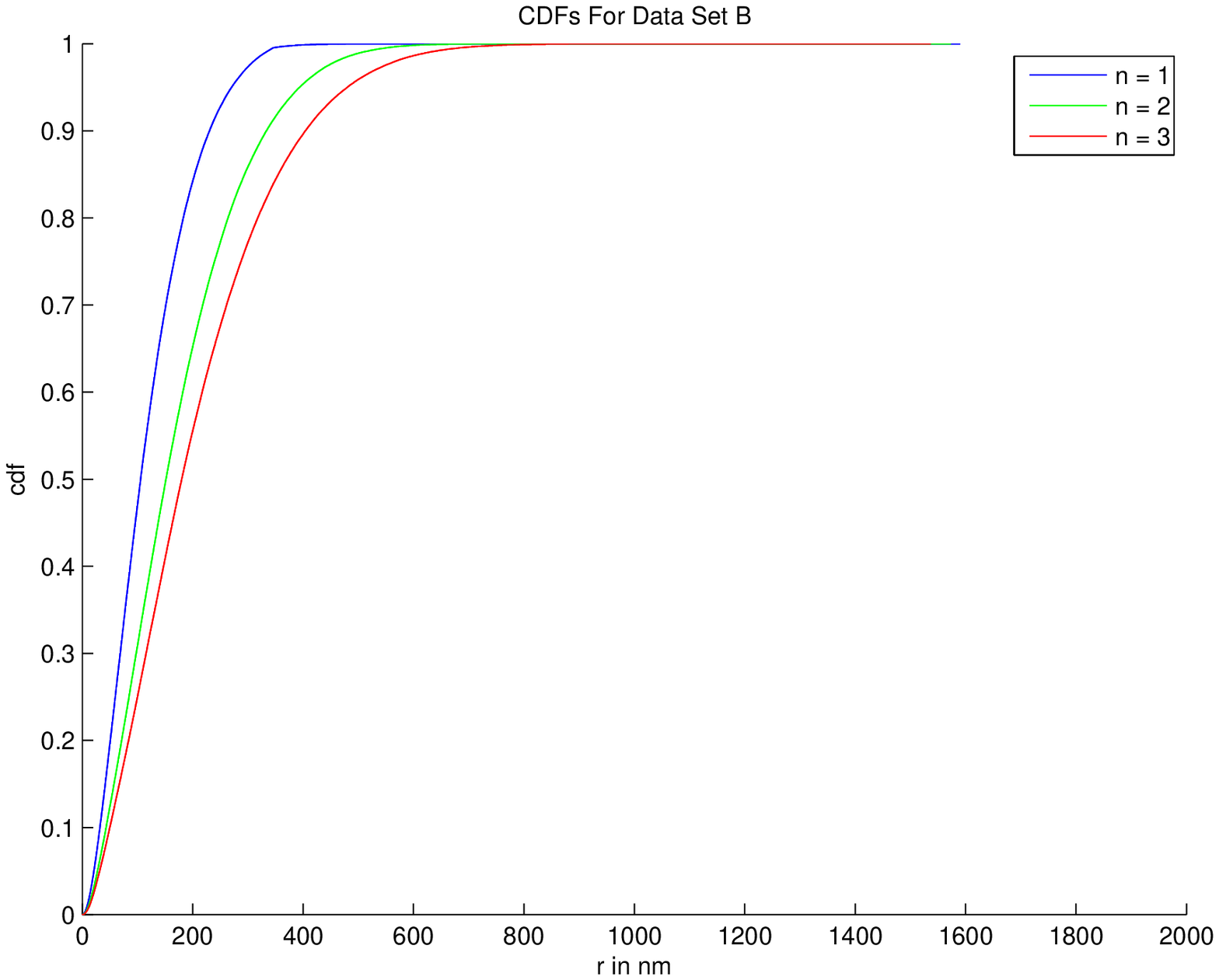}
}
\end{tabular}
\caption{The CDFs for data sets A and B} 
\label{CDFs}
\end{figure}

\begin{figure}[ht]
\centering
\begin{tabular}{cc}
\subfloat[Data set A]
{
\includegraphics[width=0.4\textwidth]{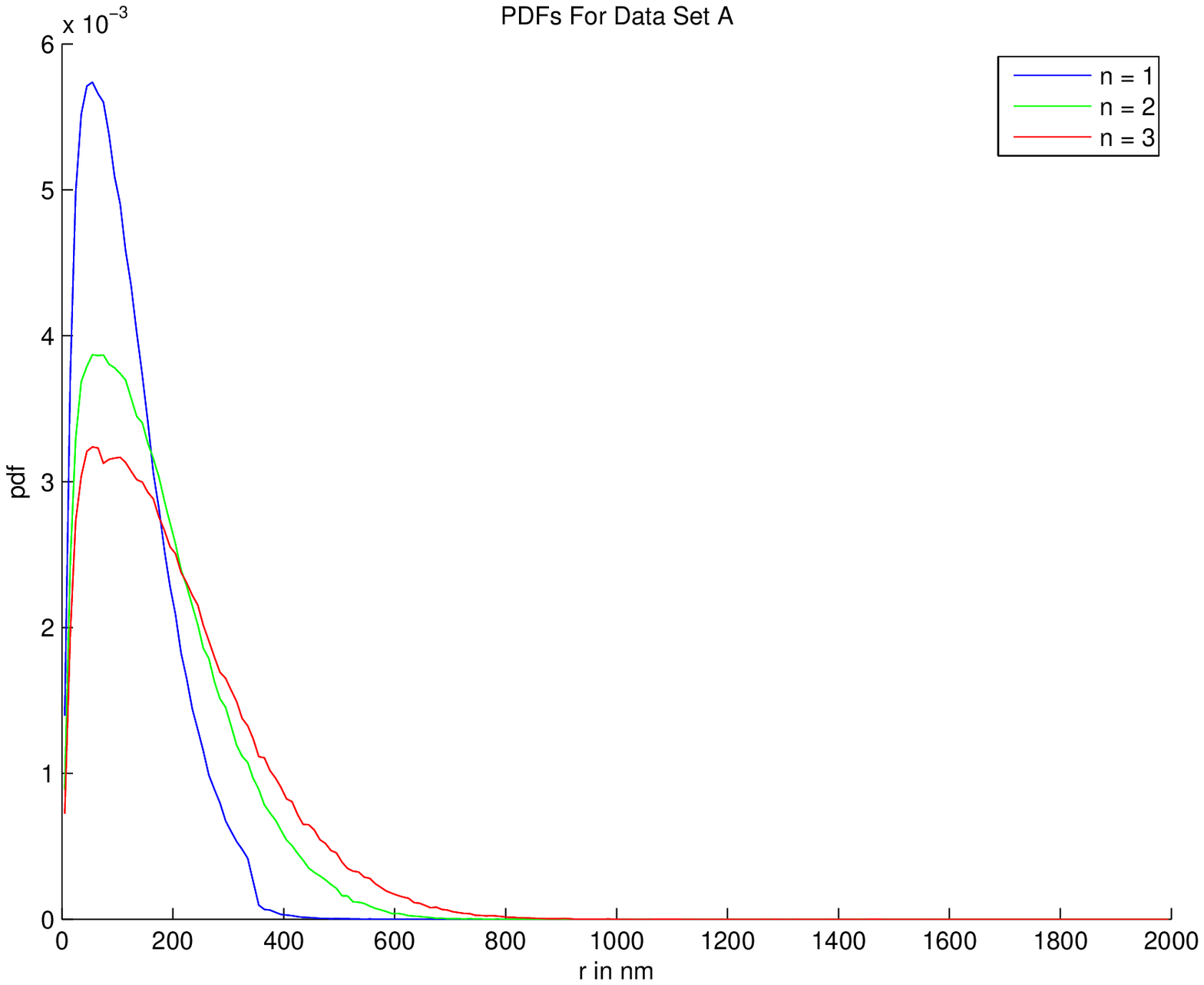}
} \qquad
&
\subfloat[Data set B]
{
\includegraphics[width=0.4\textwidth]{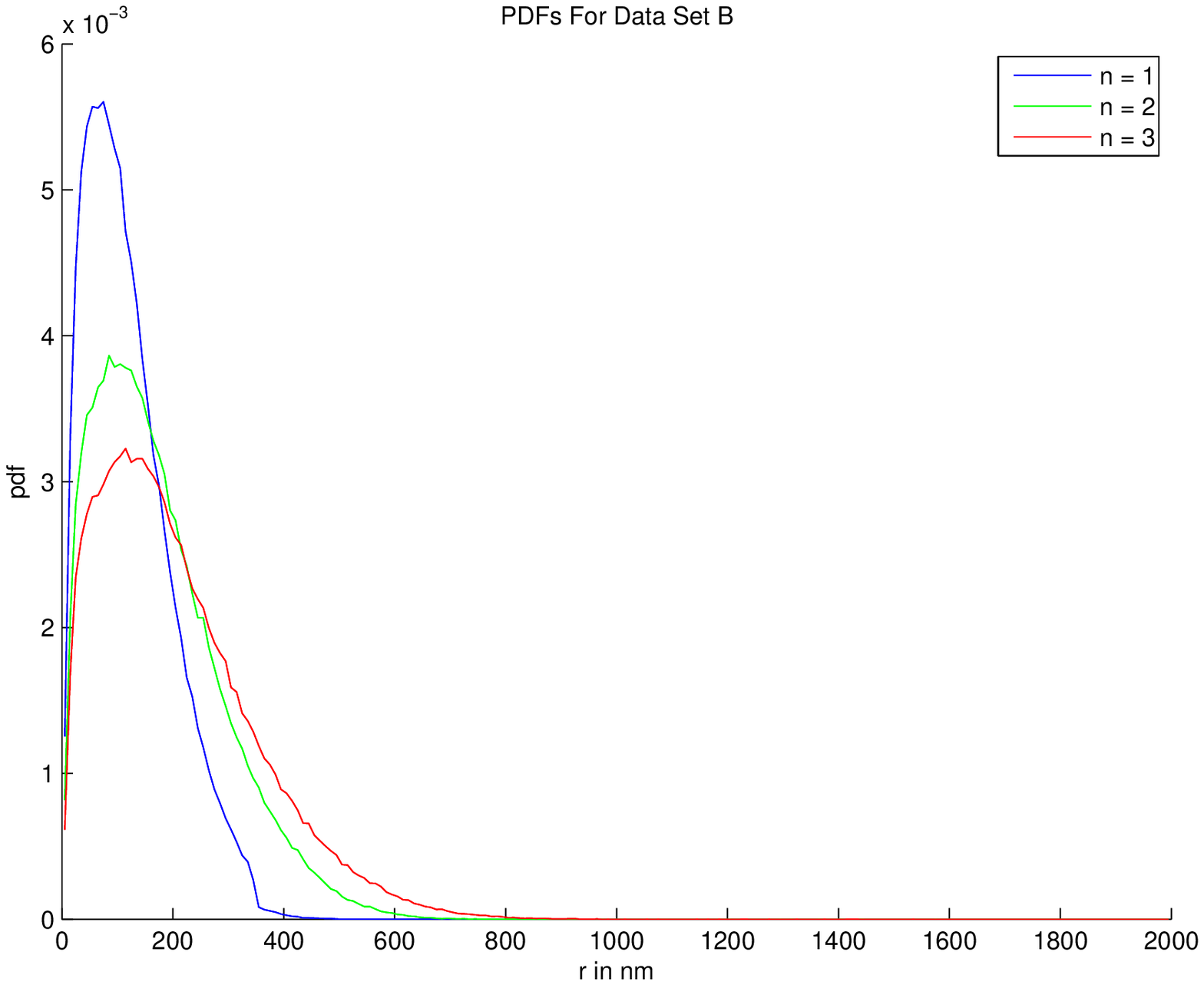}
}
\end{tabular}
\caption{The PDFs for data sets A and B} 
\label{PDFs}
\end{figure}

Our first step in analyzing the data is to plot the cumulative distribution
functions (CDFs) and probability distribution functions (PDFs) of the jump
lengths which are shown in Figures \ref{CDFs} and \ref{PDFs}.
The CDF is computed by first sorting the lengths into increasing
size.  Assuming that there are $I>0$ lengths $r_i$,
$1 \leq i \leq I$ then increasing size means that
$r_i \leq r_{i+1}$, $1 \leq i \leq I -1$.  The CDFs shown in
Figure \ref{CDFs} are determined by the pairs $(r_i,i/I)$.
This method of determining the CDFs is helpful as it makes use
of all of the data without any averaging like that which occurs
when binning the data.  Next we find the PDFs shown in Figure
\ref{PDFs} by the standard method of binning the data using 200 bins,
thus averaging over about 1/2\% of the data for each value of the PDF.
These plots use all of the valid jumps in each of the data sets and
$n$ is the number of time steps determining the jumps.
Observe that as the number of time steps increase, the height of the
PDFs decreases while the width increases, suggesting that the PDFs
are related by a scaling.  As can be seen in Figure \ref{PDFs} and noted
in \cite{espinoza12}, for the single time step jumps ($n=1$),
the data for $r> 346\nm$ have significant errors and consequently are not
used in our analysis. As we are most interested in short jumps, this
does not cause a problem.

\subsection{Scaling the Data}

\begin{figure}[ht]
\centering
\begin{tabular}{cc}
\subfloat[Data set A]
{
\includegraphics[width=0.4\textwidth]{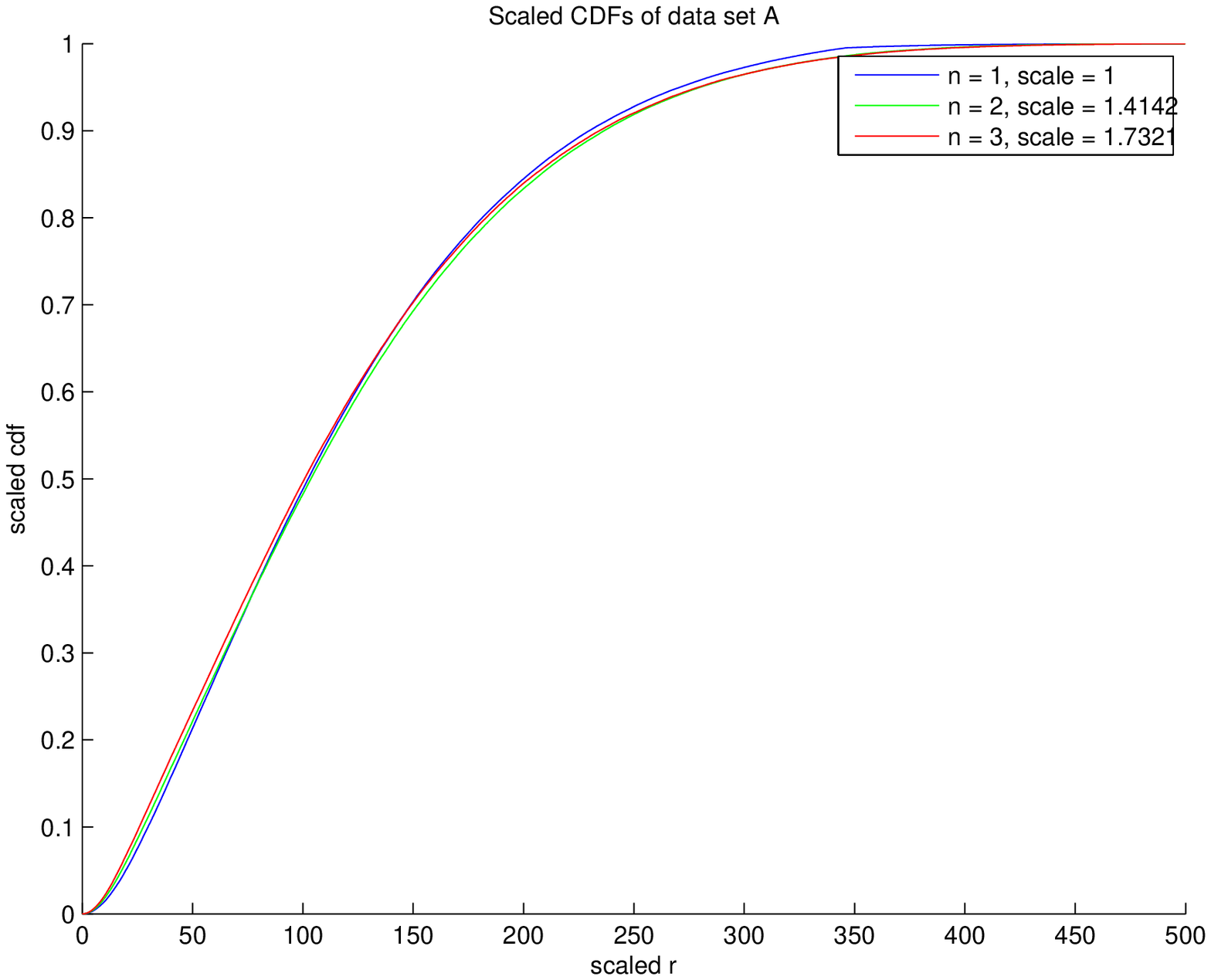}
} \qquad
&
\subfloat[Data set B]
{
\includegraphics[width=0.4\textwidth]{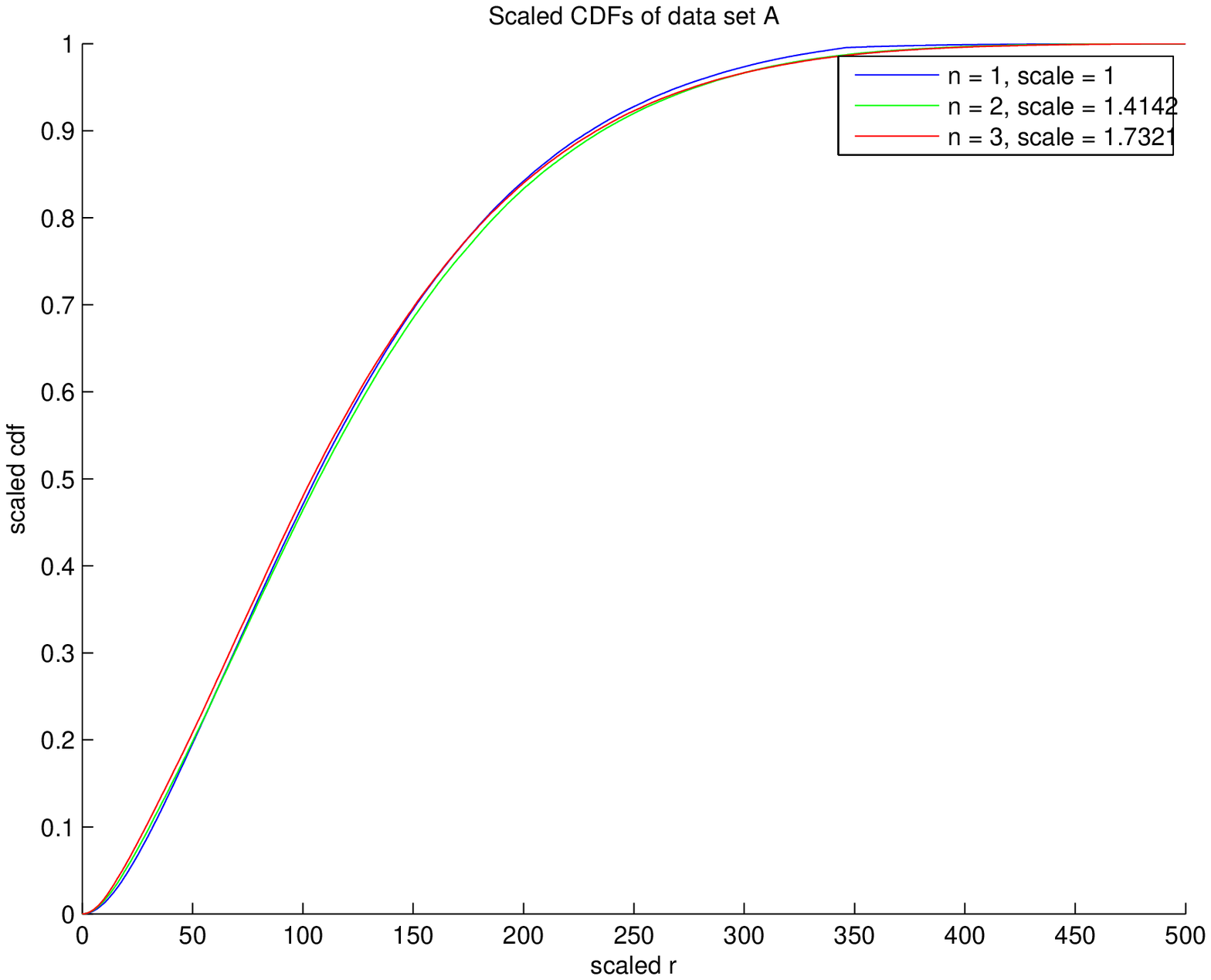}
}
\end{tabular}
\caption{The scaled CDFs for all data in the data sets A and B} 
\label{ScaledCDFs}
\end{figure}

\begin{figure}[ht]
\centering
\begin{tabular}{cc}
\subfloat[Data set A]
{
\includegraphics[width=0.4\textwidth]{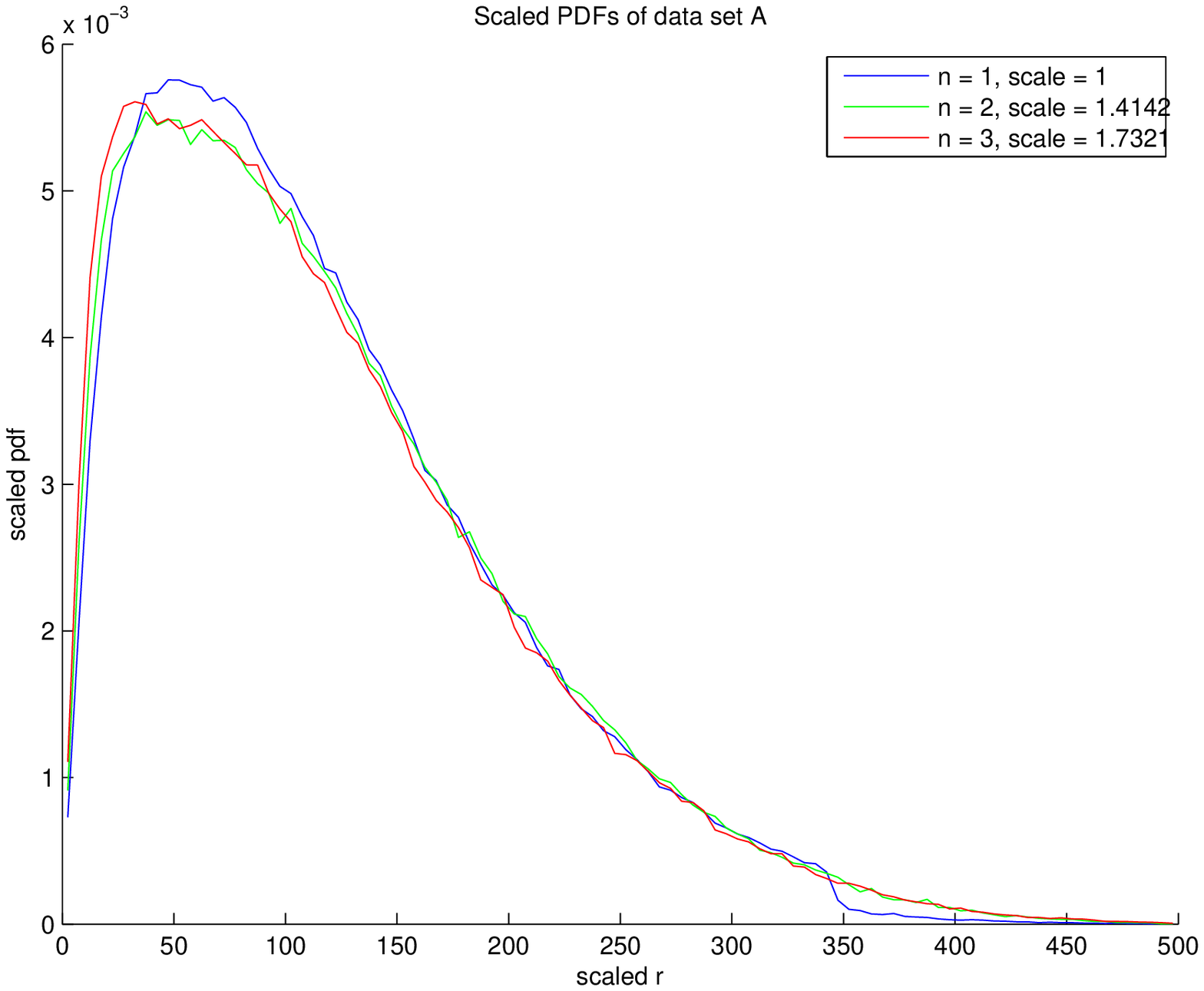}
} \qquad
&
\subfloat[Data set B]
{
\includegraphics[width=0.4\textwidth]{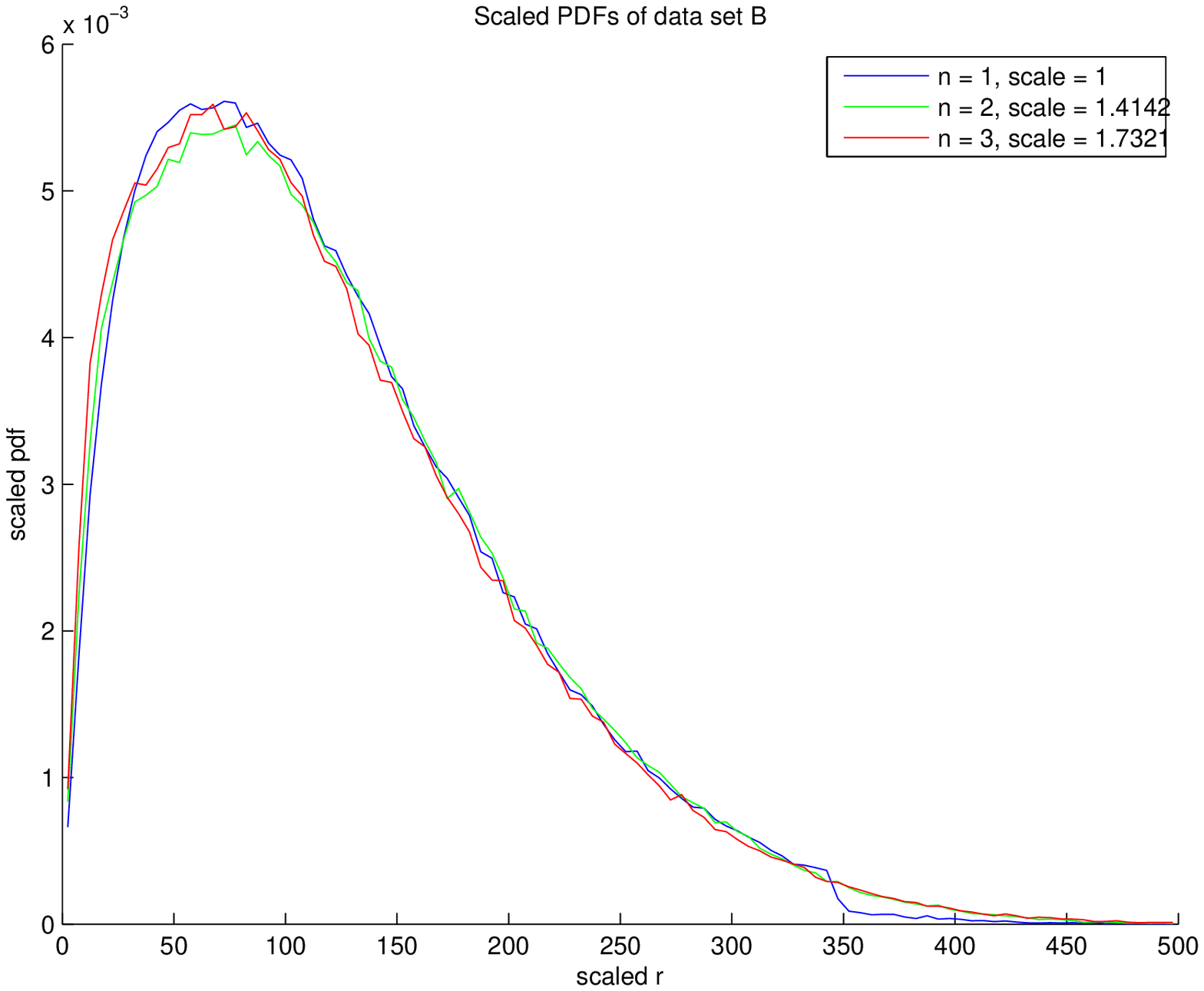}
}
\end{tabular}
\caption{The scaled PDFs for all data in the data sets A and B} 
\label{ScaledPDFs}
\end{figure}

If the CDF and PDF are smooth then they are related by
\[
p(r)=\frac{d P(r)}{d r}\,.
\]
If $s>0$ is to be used as a scaling factor,
then the scaled CDF and PDF are given by
\[
P\left(\frac{r}{s}\right) \text{ and }
\frac{1}{s} \, p\left(\frac{r}{s}\right) \,.
\]


We tried scaling the data for 1, 2 and 3 time step jumps by
$\sqrt{1} = 1$, $\sqrt{2}$ and $\sqrt{3}$ which is done by
dividing the jumps by the scale factor.
The results of the scaling are shown in Figures \ref{ScaledCDFs}
and \ref{ScaledPDFs}.  It is a surprise that the scaled CDFs and PDFs are
so similar and the scaling property is the same as for Brownian motion.
Thus, even though the components of the jumps are not normally distributed,
the angles of the jumps can be modeled as uniformly distributed and the
jump lengths can be modeled by a single PDF or CDF that is scaled by
$\sqrt{t}$.  We now turn to quantifying this idea.

Because of the errors in the single time step jumps for for $r> 346\nm$
we only analyze the scaled data for $r \leq 346\nm$ which corresponds
to the unscaled jumps satisfying
$r \leq 346\nano\meter$ when $n = 1$,
$r \leq 489\nano\meter$ when $n = 2$ and
$r \leq 599\nano\meter$ when $n = 3$.
However, we plot our results for scaled data satisfying
$r \leq 500\nano\meter$ which scales to
$r \leq 500\nano\meter$ when $n = 1$,
$r \leq 707\nano\meter$ when $n = 2$ and
$r \leq 866\nano\meter$ when $n = 3$
to emphasize that are results are quite good even for large jumps

\newpage \clearpage
\setcounter{equation}{0} \setcounter{figure}{0} \setcounter{table}{0}
\section{Fitting the Data}

We begin by describing the double power law distribution and
it's scaling law. We then show how to generate random numbers
having the double power law probability distribution function
and finish by using the double power law to fit the jump data.

\subsection{The double power law distribution}

In order for the parameters to have a more intuitive meaning we will
change the notation from that in \cite{espinoza12}.  The
functional form is chosen in a way that the PDF $p(r)$ is a power
law for both small and large $r$, which is substantially more
restrictive than having only one of these conditions hold.
Additionally, we required that the PDF has a simple CDF so that
it is easy to simulate random numbers with the given PDF.
The result is our {\em double power law}.

For small $r \geq 0$, we require
\[
p(r) \approx C_1 \, r^{\alpha - 1} \,,\quad \alpha \geq 1 \,,
\]
because then $\alpha$ corresponds to the dimension of the
space (see \eqref{Chi}) in which the diffusion is occurring.
For large $r$ we want the PDF to decay rapidly, so we choose 
\[
p(r) \approx \frac{C_2}{r^\beta} \,,\quad \beta > 1 \,.
\]
The condition on $\beta$ guarantees that the PDF has a finite
integral.  We find a good choice for the PDF $p$ and corresponding
CDF $P$ is
\[
P(r) = 1 - \frac{1}{(1 + r^\alpha)^{(\beta-1)/\alpha}} \,,\quad
p(r) =
\frac{(\beta-1) \, r^{\alpha-1}}{(1 + r^\alpha)^{(\alpha-1+\beta)/\alpha}}
\,,\quad r \geq 0 \,.
\]

Also, to make the values of the parameters easy to interpret, we need to
find a scale factor $S$ so that the second moment of the scaled power law
will be one.  The first and second moments of any distribution $p$ are:
\[
m_1 = \int_{0}^{\infty} r \, p(r) \, d r \,,\quad
m_2 = \int_{0}^{\infty} r^2 \, p(r) \, d r \,.
\]
For the power law distribution, integration gives
\[
m_1 =
\frac{\Gamma \left(\frac{\alpha+1}{\alpha }\right)
	\Gamma \left(\frac{\beta -2}{\alpha}\right)}
	{\Gamma \left(\frac{\beta -1}{\alpha}\right)}
\,,\quad
m_2 = 
\frac{\Gamma \left(\frac{\alpha+2}{\alpha }\right)
	\Gamma \left(\frac{\beta -3}{\alpha }\right)}
	{\Gamma \left(\frac{\beta-1}{\alpha }\right)} \,.
\]

If we set
\begin{equation}\label{PLPDFCDF}
pl(r) = S \,\, p(S\,r) \,,\quad PL(r) = P(S \, r) \,,
\end{equation}
then
\[
\int_{0}^{\infty} pl(r) \, d r = 1 \,,
\]
and the first and second moments of this distribution are
\[
M_1 =\frac{m_1}{S} \,,\quad M_2 = \frac{m_2}{S^2} \,.
\]
Consequently, if
\begin{equation}\label{Svalue}
S = \sqrt{m_2} \,,
\end{equation}
then 
\[
M_2 = 1 \,.
\]
The required double power law distributions are then given by
\eqref{PLPDFCDF}.

\subsection{Simulating the double power law distribution}

\begin{figure}[ht]
\centering
\begin{tabular}{cc}
\subfloat[The CDF]
{
\includegraphics[width=0.5\textwidth]{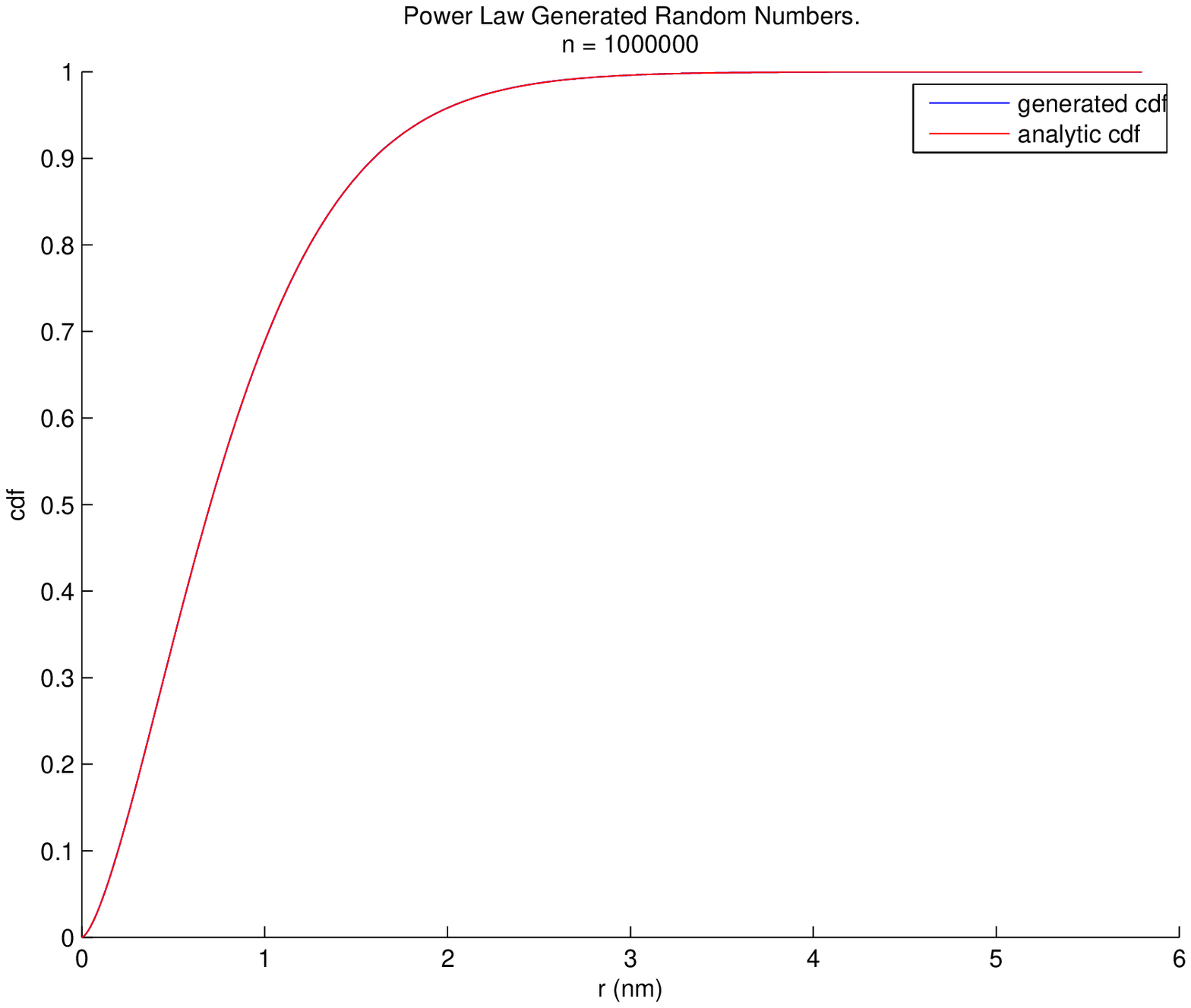}
} \qquad
&
\subfloat[The PDF]
{
\includegraphics[width=0.5\textwidth]{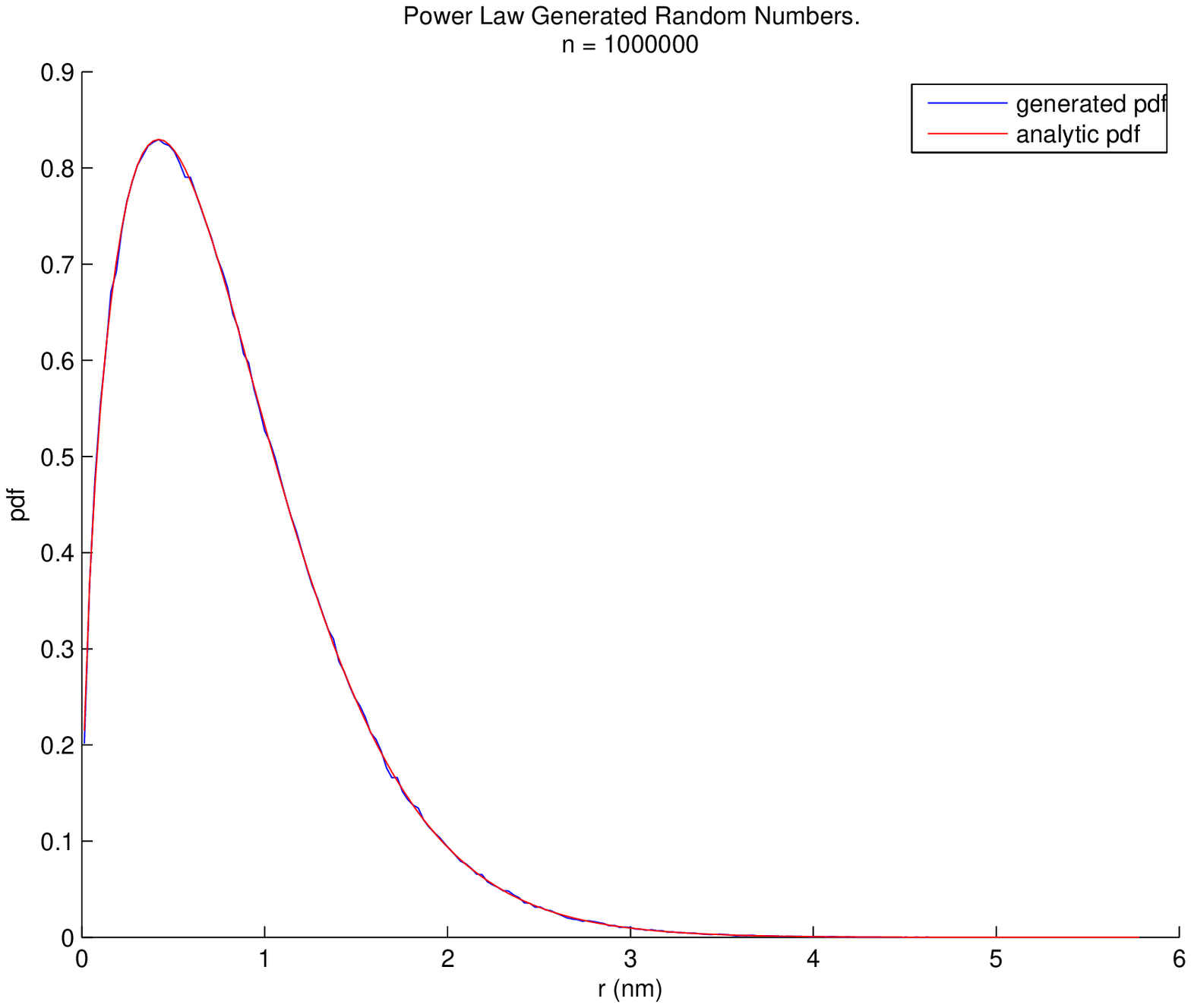}
}
\end{tabular}
\caption{The analytic and simulated $pl$ and $PL$ distributions
with $\sigma = 1$ are essentially identical.}
\label{GeneratedPL}
\end{figure}

As is well known, if $r$ is a uniformly distributed random number in
$[0,1]$, then solving $P(r) = u$ will produce random number with the
given CDF $P$ and associated PDF $p$.  Therefore, we can generate
random number with the $pl$ distribution by solving
\[
PL(r ) = u
\]
where $u$ is uniformly distributed. This gives
\[
r =  \frac{1}{S}
\left(-1 + (1 - u)^{-\alpha/(-1 + \beta)}\right)^{1/\alpha} \,,
\]
where $S$ is given in \eqref{Svalue}.
Because $u$ is a uniformly distributed random number in $[0,1]$, then
so in $1-u$ so this can be simplified to
\[
 r =  \frac{1}{S} \, \left(-1 + u^{-\alpha/(-1 + \beta)}\right)^{1/\alpha} \,.
\]
We numerically checked that this distribution has $\sigma = 1$ and
plot the comparisons of the simulated and analytic PDF and CDF in
Figure \ref{GeneratedPL}.  Double power law random numbers with second
moment $\sigma^2$ are given by $ \sigma \, r $.

\subsection{Fitting the PDF and CDF}

\begin{figure}[ht]
\centering
\begin{tabular}{cc}
\subfloat[CDF]
{
\includegraphics[width=0.4\textwidth]{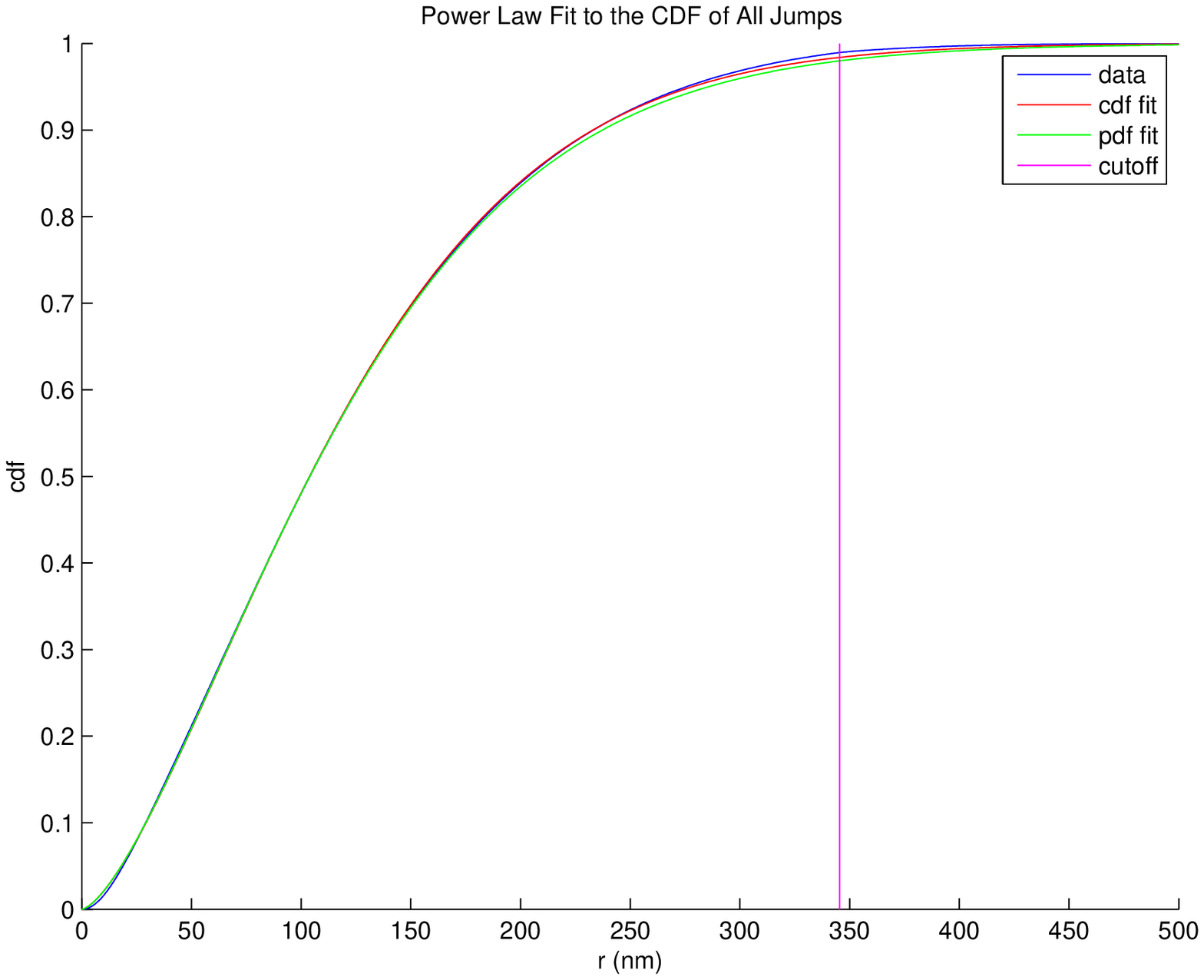}
} \qquad

&
\subfloat[PDF]
{
\includegraphics[width=0.4\textwidth]{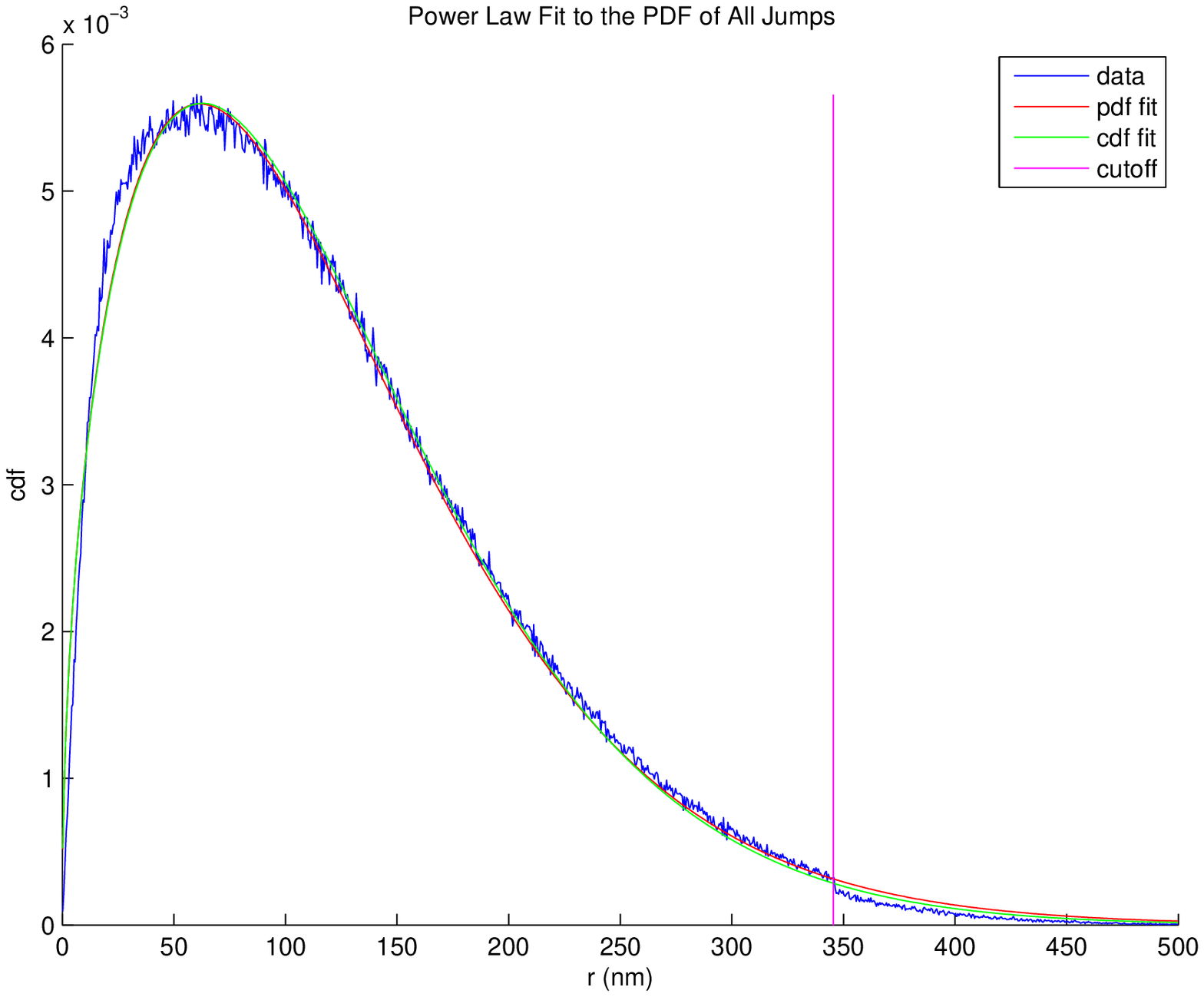}
}
\end{tabular}
\caption{Plots of the fits of both the PDF and CDF of the data.
The vertical red line indicates the cut off used for fitting the data.} 
\label{PLFitCDFPDF}
\end{figure}

We now use least-squares data fitting to find a double exponential
power law that fits the jump data, that is, we must estimate the
parameters $\alpha$ and $\beta$ and the scale parameter $s$.
To obtain maximal accuracy for the parameters, we combine all the
scaled jumps sizes from data sets $A$ and $B$ and scaled jumps sizes
for 1, 2 and 3 time steps. As noted above, and can be seen in Figure
\ref{PLFitCDFPDF}b, the data for $r> 346\nm$ have significant errors,
so we only fit jumps for  $r \leq 346\nm$ which gives 1,971,537 values
that can be used for the fits, which are excellent for $r \leq 346\nm$.

\begin{table}
\begin{center}
 \begin{tabular}{|r|rrr|r|}
\hline
    & $\alpha$ & $\beta$ & $s(\nm)$ & residual \\
\hline
PDF & 1.49 &  39.47 & 148.52 & 1.9e-08 \\
CDF & 1.49 & 256.86 & 145.73 & 6.8e-06 \\
\hline
\end{tabular}
\caption{The values $\alpha$, $\beta$, $s$ and the residual for the fits.
\label{Parameters}}
\end{center}
\end{table}

In fact, we fit both the PDF and CDF independently to obtain
\begin{align*}
& \alpha_\text{CDF} \,,\quad \beta_\text{CDF} \,,\quad s_\text{CDF} \,, \\
& \alpha_\text{PDF} \,,\quad \beta_\text{PDF} \,,\quad s_\text{PDF} \,.
\end{align*}
But the parameters of the fit of the PDF can be used in the fit to the
CDF and conversely.  As is common with fitting problems, the solution
is not unique, as shown in Table \ref{Parameters}, but the graphs of the
fits are essentially the same. We consider the fit with $\beta \approx 40$
the simplest and thus the better fit.  We plot both of these fits in 
Figure \ref{PLFitCDFPDF}.  We are most interested in the
parameter $\alpha$ which describes the short jumps and is {\em unique}.
Also, the square root of the second moment of all of the scaled jumps
is $143.85\nm$ and indeed the scale factors, as expected, are close to
this this value.

\newpage \clearpage
\setcounter{equation}{0} \setcounter{figure}{0} \setcounter{table}{0}
\section{Results and Discussion}

We have shown that the small spatial and short time motion of
QDs labeling proteins on the cell membrane can be characterized by a single
double power law and square root of time scaling. In fact, this
distribution is a good fit for all of the scaled jumps under $340 \nm$.
This power law quantifies the idea that this motion has
substantially more short jumps than if the motion was a
Brownian random walk.  Another way to say this is that the motion
can be viewed as diffusion in a space of dimension 3/2, which
also quantifies the small scale restrictions of the motion.
We note that this notion of diffusion seems not to be closely
related to diffusion on fractals \cite{prehl10} where the MSD
was used to characterize the motion.  On the other hand, the fact
that the jumps over a few time steps are all described by one
probability distribution and a scale factor suggests that this
distribution is {\em stable} \cite{nolan13} and thus captures
important properties of the motion.

To accuratly study the interaction of molecules on a cell membrane
one can use a stochastic simulator where the time step is chosen so
the the proteins typically move only a few nanometers. Our results
can be extend to provide the probabilty that two proteins will interact
in time intervals where the proteins move a few hundred nanometers,
providing a tool for significantly reducing the simulation cost.

We note that the Central Limit Theorem puts a strong restrictions on
modeling the motion.  To confirm this, we used our double power law
random number generator combined with uniformly generated angles to
simulate random walks.  For a 100 walkers going 10 time steps,
the distribution of the distance moved over the 10 steps is now
normally distributed as predicted by the Central Limit Theorem.
This implies that the motion cannot be accurately modeled as a IID
random walk in a homogeneous medium.

Researchers associated with the New Mexico Center for the Spatiotemporal
Modeling of Cell Signaling have created models of the cell membrane
that include models of actin filament barriers and lipid rafts where
the jump data has properties similar to the jump data analyzed here. 
It is hoped that these models will provide a connection between the
statistics of the membrane structure and the resulting PDF of the jump sizes.

Our results complement recent results on the analysis of the motion
of proteins on longer time scales using mean squared displacement ideas.
For example:
\cite{arnspangscwl2013} uses k-space image correlation spectroscopy analysis
to study single molecule density data of lipids and proteins labeled
with quantum dots;
\cite{perssonlue2013} uses variational Bayesian treatment of hidden
Markov models to analyze many short tracks identifying several diffusion states;
\cite{marquezlb2012} uses nonergodic motion, fractal structures,
and multifractional random motion to study the motion of proteins;
\cite{monniergmhlb2012} uses Basian statistics to better
quantify noise from sampling limitations and biological heterogeneity;
\cite{weigelstk2011} shows that both ergodic and a nonergodic
process exist in the plasma membrane and that the ergodic process
resembles a fractal structure originating in the  macromolecular
crowding in the cell membrane;
\cite{dascc2009} uses hidden Markov models are used to identify
multiple states of diffusion within experimental trajectories.


\newpage \clearpage
\bibliography{citations}

\newpage \clearpage
\begin{appendix}
\newpage \clearpage
\setcounter{equation}{0} \setcounter{figure}{0} \setcounter{table}{0}
\section{The fits of the CDFs and PDFs \label{Fits Individual Data}}

In this section we show how well the computed PDF power laws
fit the unscaled data for 1, 2 and 3 jumps and data sets $A$ and $B$.
Again we see that the fits are excellent, but now the noise
in the data is more apparent. 

\begin{figure}[ht]
\centering
\begin{tabular}{cccc}
\subfloat[1 jump, data set A]
{
\includegraphics[width=0.30\textwidth]{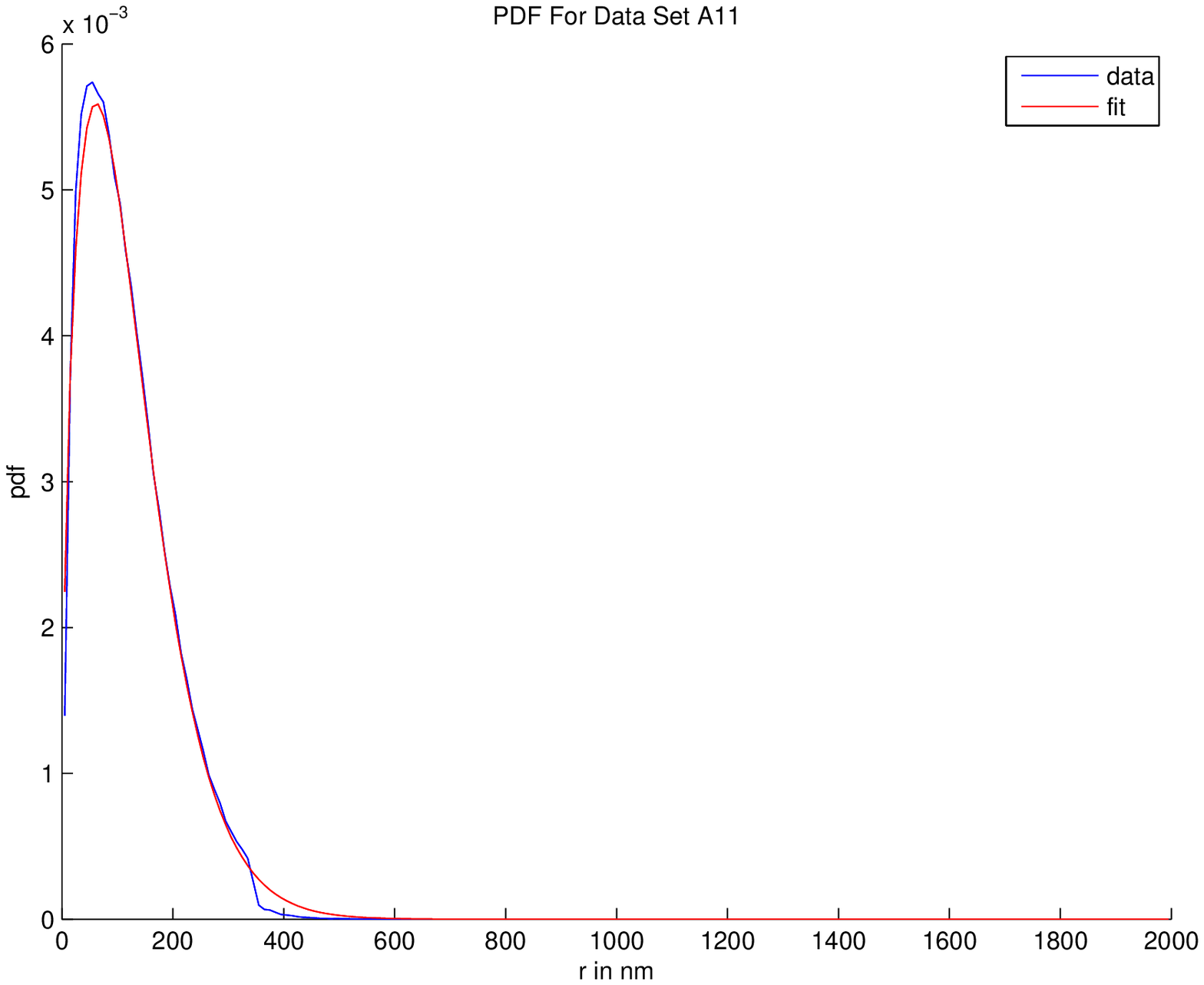}
} \qquad
&
\subfloat[1 jump, data set B]
{
\includegraphics[width=0.30\textwidth]{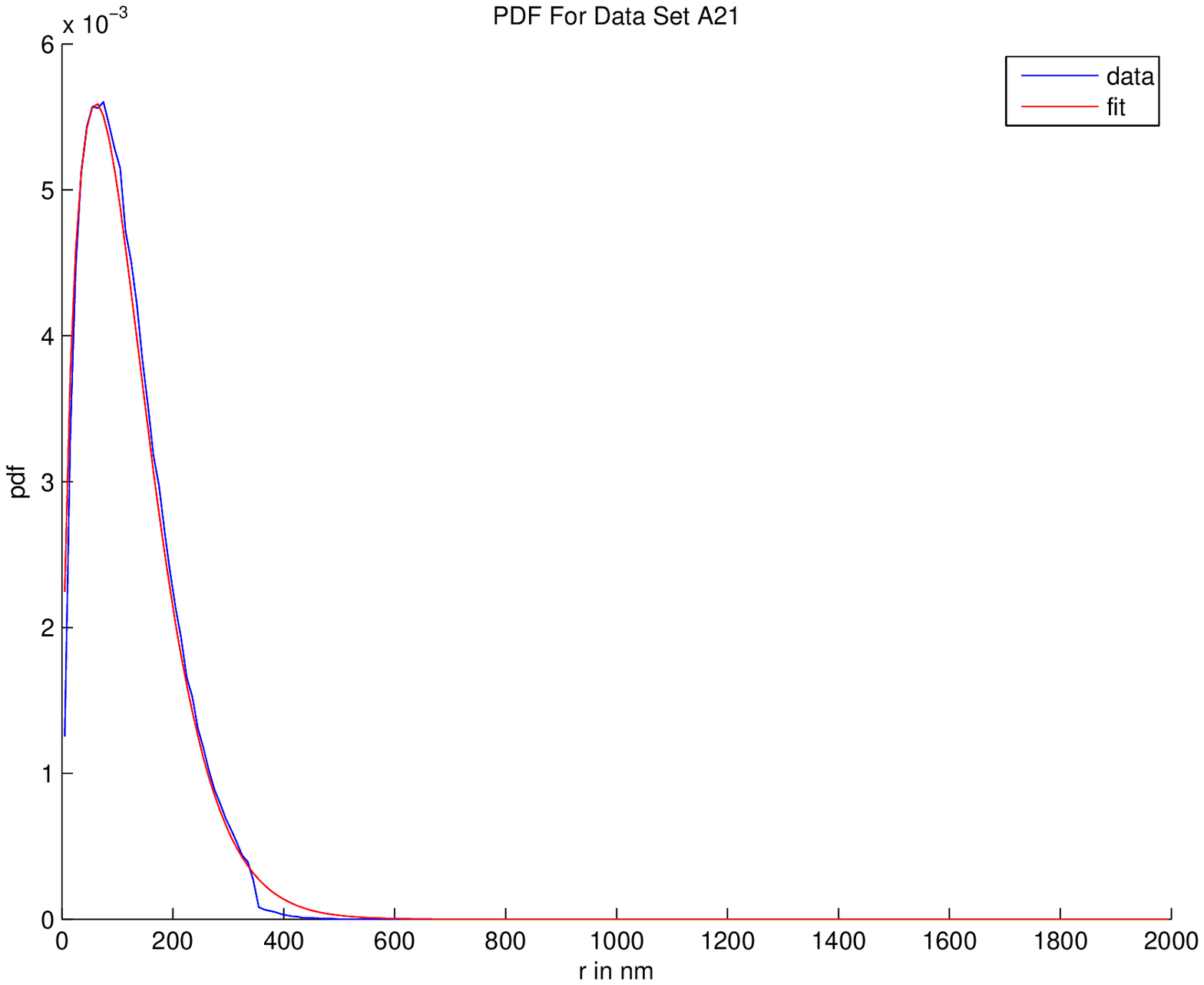}
} \\
\subfloat[2 jumps, data set A]
{
\includegraphics[width=0.30\textwidth]{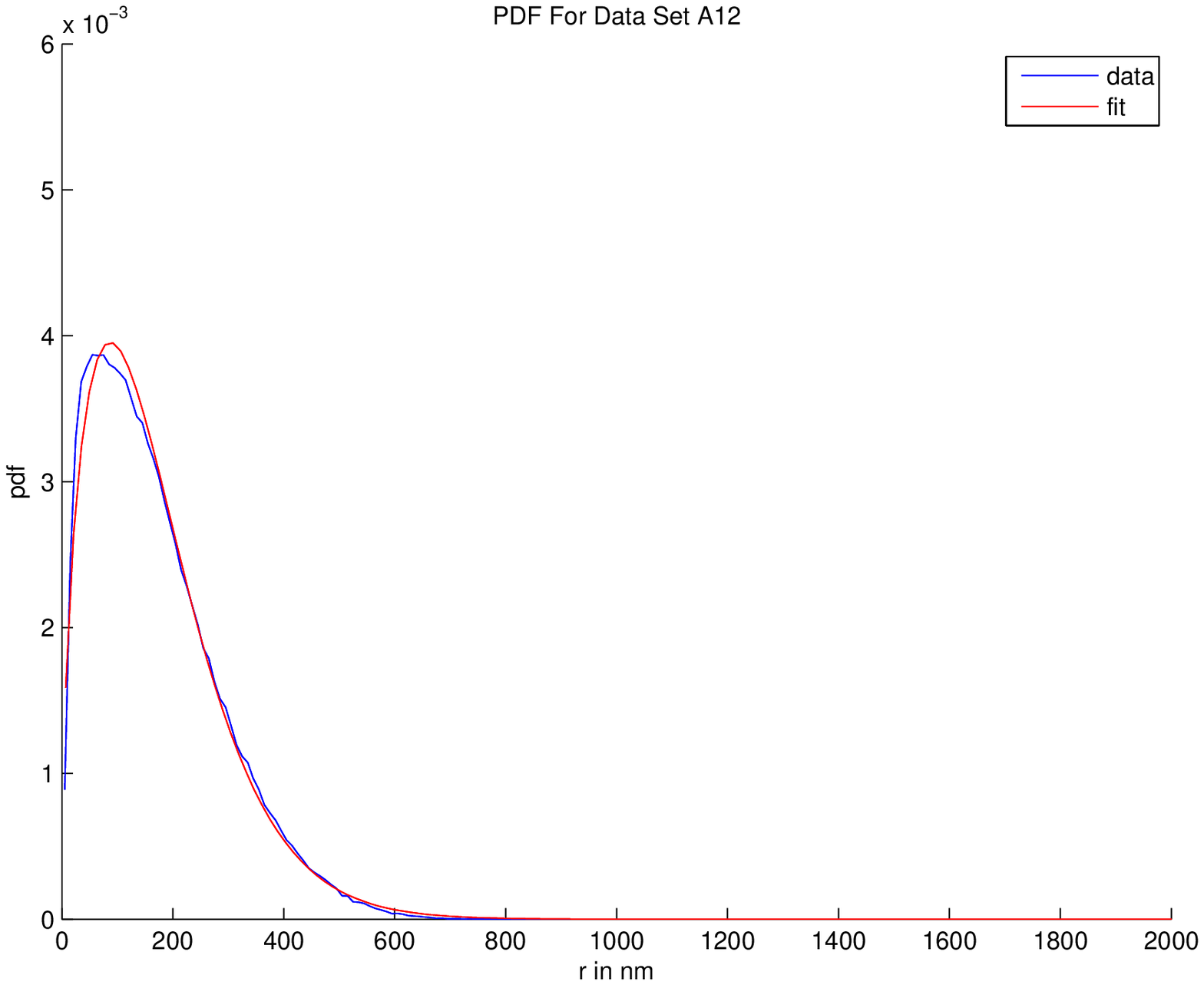}
} \qquad
&
\subfloat[2 jumps, data set B]
{
\includegraphics[width=0.30\textwidth]{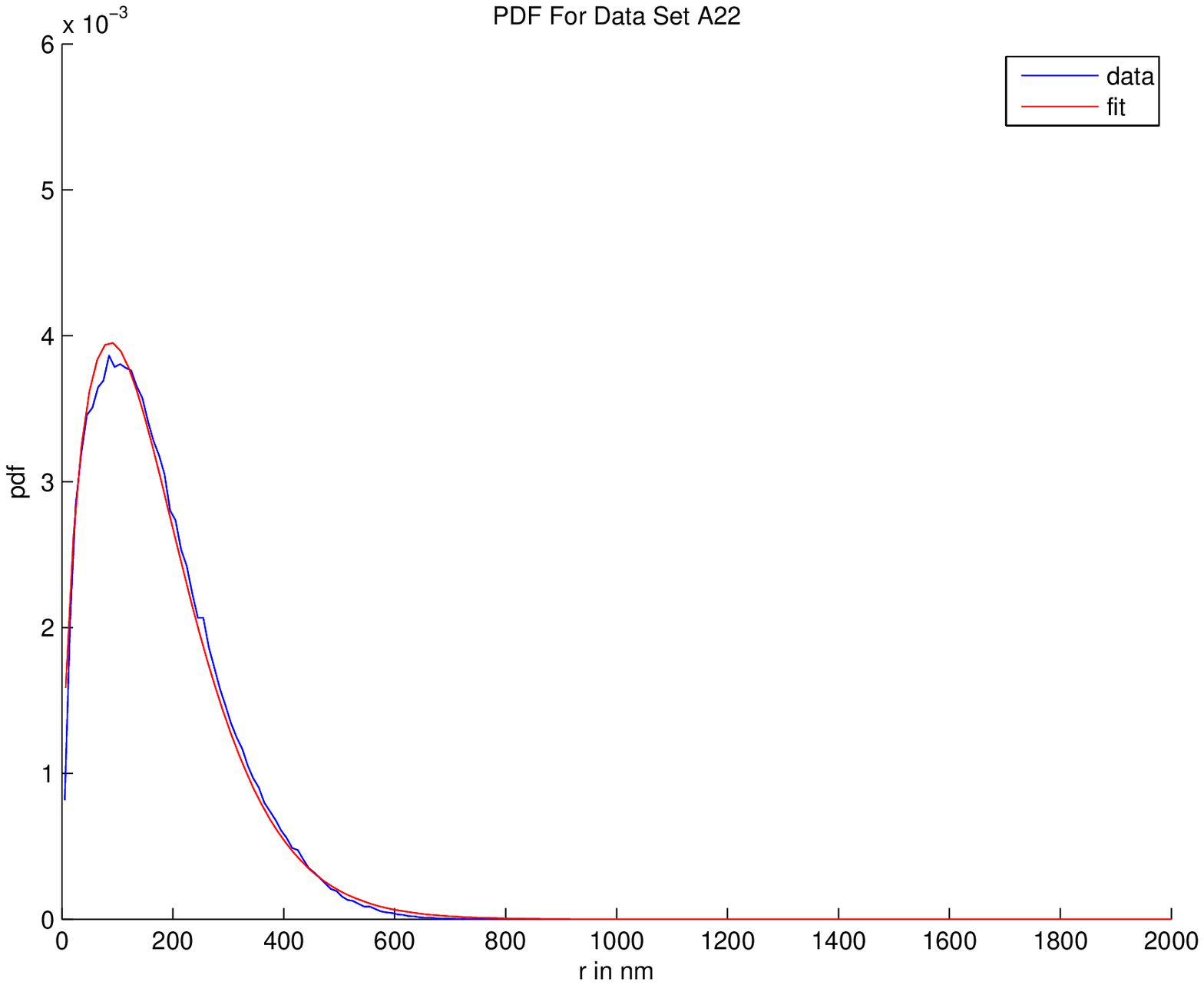}
} \\
\subfloat[3 jumps, data set A]
{
\includegraphics[width=0.30\textwidth]{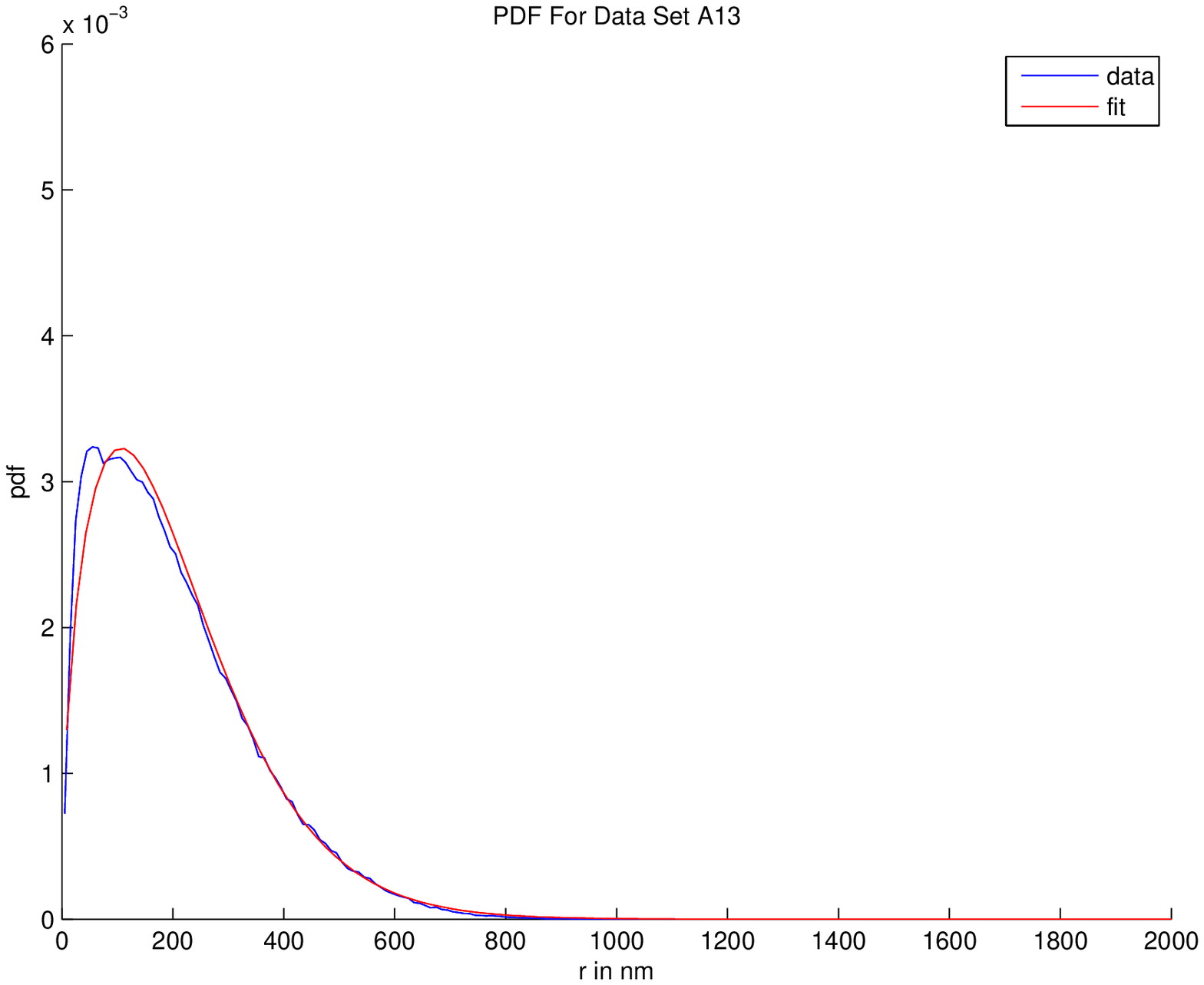}
} \qquad
&
\subfloat[3 jumps, data set B]
{
\includegraphics[width=0.35\textwidth]{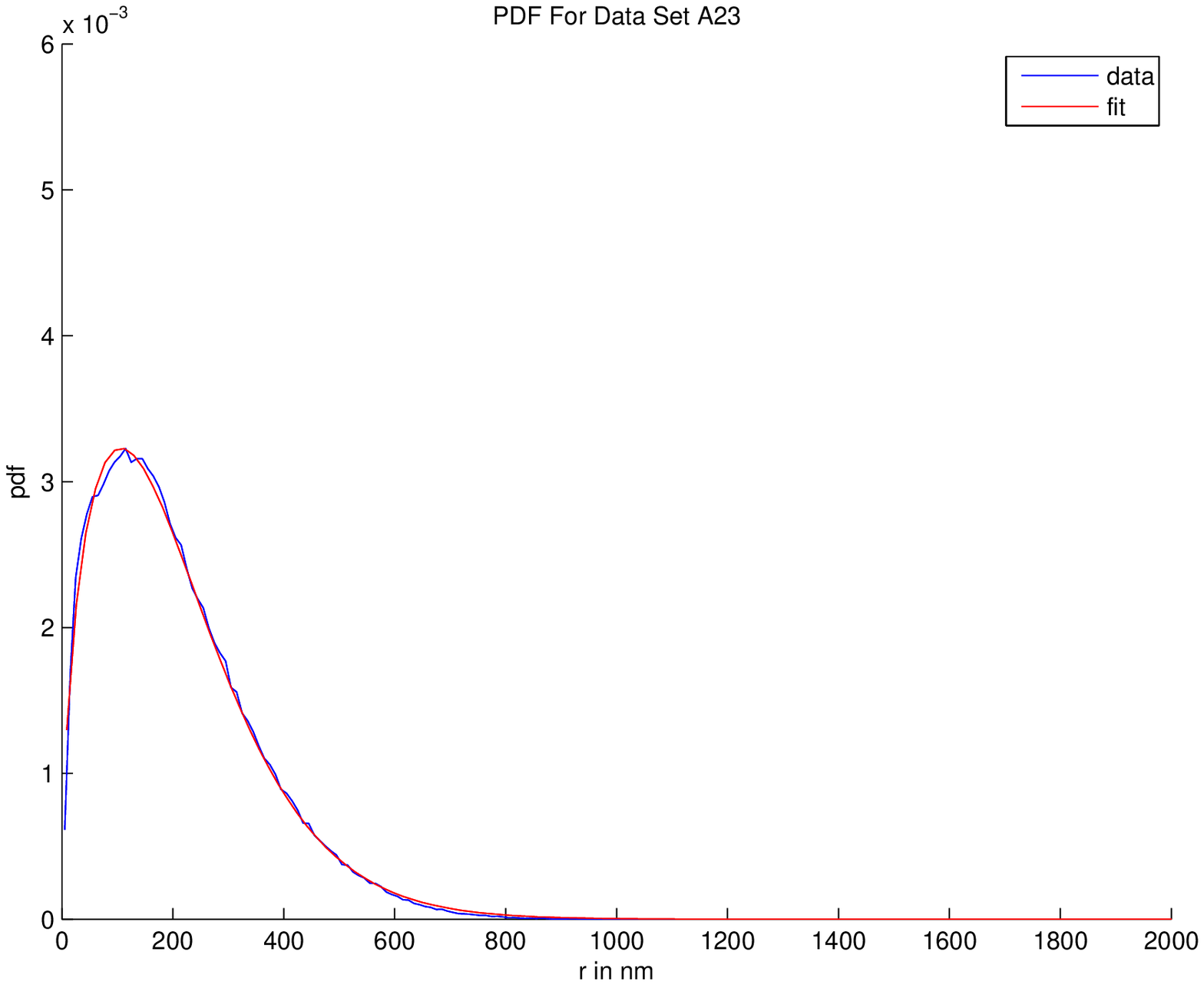}
}
\end{tabular}

\caption{The PDFs fits for data sets A and B and 1, 2, and 3 jumps.} 
\end{figure}

\end{appendix}

\newpage \clearpage
\end{document}